\documentclass[conference]{IEEEtran}
\usepackage{lmodern}
\usepackage{amssymb,amsmath}
\usepackage{ifxetex,ifluatex}
\usepackage{fixltx2e} % provides \textsubscript
\ifnum 0\ifxetex 1\fi\ifluatex 1\fi=0 % if pdftex
  \usepackage[T1]{fontenc}
  \usepackage[utf8]{inputenc}
\else % if luatex or xelatex
  \ifxetex
    \usepackage{mathspec}
  \else
    \usepackage{fontspec}
  \fi
  \defaultfontfeatures{Ligatures=TeX,Scale=MatchLowercase}
\fi
\IfFileExists{upquote.sty}{\usepackage{upquote}}{}
\IfFileExists{microtype.sty}{%
\usepackage{microtype}
\UseMicrotypeSet[protrusion]{basicmath} % disable protrusion for tt fonts
}{}
\usepackage[unicode=true]{hyperref}
\hypersetup{
            pdftitle={AIWC: OpenCL-based Architecture-Independent Workload Characterization},
            pdfkeywords={workload characterization, benchmarking, HPC},
            pdfborder={0 0 0},
            breaklinks=true}
\IfFileExists{parskip.sty}{%
\usepackage{parskip}
}{% else
\setlength{\parindent}{0pt}
\setlength{\parskip}{6pt plus 2pt minus 1pt}
}
\providecommand{\tightlist}{%
  \setlength{\itemsep}{0pt}\setlength{\parskip}{0pt}}
\ifx\paragraph\undefined\else
\let\oldparagraph\paragraph
\renewcommand{\paragraph}[1]{\oldparagraph{#1}\mbox{}}
\fi
\ifx\subparagraph\undefined\else
\let\oldsubparagraph\subparagraph
\renewcommand{\subparagraph}[1]{\oldsubparagraph{#1}\mbox{}}
\fi

\makeatletter
\def\fps@figure{htbp}
\makeatother

\usepackage{booktabs}
\usepackage[figurename={Figure.},tablename={Table.},listfigurename={List of Figures.},listtablename={List of Tables.}]{caption}
\usepackage{tabularx}
\usepackage{cite}
\usepackage{amsmath,amssymb,amsfonts}
\usepackage{textcomp}
\usepackage{xargs}
\usepackage[colorinlistoftodos,prependcaption,textsize=small,color=yellow]{todonotes}
\usepackage{regexpatch}
\usepackage{adjustbox}
\usepackage{etoolbox}
\usepackage{listings}
\usepackage{multicol}
\usepackage[noend,ruled]{algorithm2e}
\usepackage{float}
\usepackage{threeparttable}
\usepackage[binary-units=true]{siunitx}
\def\BibTeX{{\rm B\kern-.05em{\sc i\kern-.025em b}\kern-.08em T\kern-.1667em\lower.7ex\hbox{E}\kern-.125emX}}
\makeatletter
\@ifpackageloaded{subfig}{}{\usepackage{subfig}}
\@ifpackageloaded{caption}{}{\usepackage{caption}}
\AtBeginDocument{%
\renewcommand*\figurename{Figure}
\renewcommand*\tablename{Table}
}
\AtBeginDocument{%
\renewcommand*\listfigurename{List of Figures}
\renewcommand*\listtablename{List of Tables}
}
\@ifpackageloaded{float}{}{\usepackage{float}}
\floatstyle{ruled}
\@ifundefined{c@chapter}{\newfloat{codelisting}{h}{lop}}{\newfloat{codelisting}{h}{lop}[chapter]}
\floatname{codelisting}{Listing}

\makeatother

\title{AIWC: OpenCL-based Architecture-Independent Workload Characterization}
\date{}

\begin{document}
\makeatletter
\let\oldlt\longtable
\let\endoldlt\endlongtable
\def\longtable{\@ifnextchar[\longtable@i \longtable@ii}
\def\longtable@i[#1]{\begin{figure}[t]
\onecolumn
\begin{minipage}{0.5\textwidth}
\oldlt[#1]
}
\def\longtable@ii{\begin{figure}[t]
\onecolumn
\begin{minipage}{0.5\textwidth}
\oldlt
}
\def\endlongtable{\endoldlt
\end{minipage}
\twocolumn
\end{figure}}
\newcommand{\removelatexerror}{\let\@latex@error\@gobble}
\xpatchcmd{\@todo}{\setkeys{todonotes}{#1}}{\setkeys{todonotes}{inline,#1}}{}{}
\newtoggle{IEEE-BUILD}
\toggletrue{IEEE-BUILD}
\newtoggle{ACM-BUILD}
\togglefalse{ACM-BUILD}
\newtoggle{LNCS-BUILD}
\togglefalse{LNCS-BUILD}
\makeatother
\lstset{frame=tb, tabsize=4, showstringspaces=false, numbers=none, commentstyle=\color{blue}, keywordstyle=\color{green}, stringstyle=\color{red}, basicstyle=\ttfamily\footnotesize, xleftmargin=3.4pt, xrightmargin=3.4pt}
\author{\IEEEauthorblockN{Beau Johnston}
\IEEEauthorblockA{\textit{Research School of Computer Science} \\
\textit{Australian National University}\\
Canberra, Australia \\
beau.johnston@anu.edu.au}
\and
\IEEEauthorblockN{Josh Milthorpe}
\IEEEauthorblockA{\textit{Research School of Computer Science} \\
\textit{Australian National University}\\
Canberra, Australia \\
josh.milthorpe@anu.edu.au}
}
\maketitle
\begin{abstract}
Measuring performance-critical characteristics of application workloads is important both for developers, who must understand and optimize the performance of codes, as well as designers and integrators of HPC systems, who must ensure that compute architectures are suitable for the intended workloads. However, if these workload characteristics are tied to architectural features that are specific to a particular system, they may not generalize well to alternative or future systems. An architecture-independent method ensures an accurate characterization of inherent program behaviour, without bias due to architecture-dependent features that vary widely between different types of accelerators.
This work presents the first architecture-independent workload characterization framework for heterogeneous compute platforms, proposing a set of metrics determining the suitability and performance of an application on any parallel HPC architecture. The tool, AIWC, is a plugin for the open-source Oclgrind simulator. It supports parallel workloads and is capable of characterizing OpenCL codes currently in use in the supercomputing setting. AIWC simulates an OpenCL device by directly interpreting LLVM instructions, and the resulting metrics may be used for performance prediction and developer feedback to guide device-specific optimizations. An evaluation of the metrics collected over a subset of the Extended OpenDwarfs Benchmark Suite is also presented.
\end{abstract}

\iftoggle{IEEE-BUILD}{
}{}

\section{Introduction}\label{introduction}

Modern high-performance computing (HPC) systems are typically heterogeneous, with a single node comprising a traditional CPU and an accelerator such as a GPU or many-integrated-core device (MIC).
High bandwidth, low latency interconnects such as the Cray XC50 \emph{Aries}, Fujitsu Post-K \emph{Tofu} and IBM Power9 \emph{Bluelink}, support tighter integration between compute devices on a node.
Some interconnects support multiple different kinds of devices on a single node, for example, \emph{Bluelink} features both NVLink support for Nvidia GPUs and CAPI for other emerging accelerators such as DSPs, FPGAs and MICs.

The OpenCL programming framework is well-suited to such heterogeneous computing environments, as a single OpenCL code may be executed on multiple different device types.
When combined with autotuning, an OpenCL code may exhibit good performance across varied devices. {[}1{]}

Application codes differ in resource requirements, control structure and available parallelism.
Similarly, compute devices differ in number and capabilities of execution units, processing model, and available resources.
Given performance measurements for particular combinations of codes and devices, it is difficult to generalize to novel combinations.
Hardware designers and HPC integrators would benefit from accurate and systematic performance prediction, for example, in designing an HPC system, to choose a mix of accelerators that are well-suited to the expected workload.

To this end, we present the Architecture Independent Workload Characterization (AIWC) tool.
AIWC simulates the execution of OpenCL kernels to collect architecture-independent features that characterize each code, which may also be used in performance prediction.

AIWC is the first workload characterization tool to support multi-threaded or parallel workloads, which it achieves by collecting metrics that indicate both instruction and thread-level parallelism.
Exploitable coarse-grained parallelism is measured by counting the number of work-items and barriers encountered.
Instructions To Barrier (ITB) and Instructions per Thread (IPT) can be used to indicate workload irregularity or imbalance.

We demonstrate the use of AIWC to characterize a variety of codes in the Extended OpenDwarfs Benchmark Suite {[}5{]}.

\section{Related Work}\label{related-work}

Oclgrind is an OpenCL device simulator developed by Price and McIntosh-Smith {[}6{]} capable of performing simulated kernel execution.
It operates on a restricted LLVM IR known as Standard Portable Intermediate Representation (SPIR) {[}7{]}, thereby simulating OpenCL kernel code in a hardware agnostic manner.
This architecture independence allows the tool to uncover many portability issues when migrating OpenCL code between devices.
Additionally, Oclgrind comes with a set of tools to detect runtime API errors, race conditions and invalid memory accesses, and generate instruction histograms.
AIWC is added as a tool to Oclgrind and leverages its ability to simulate OpenCL device execution using LLVM IR codes.

AIWC relies on the selection of the instruction set architecture (ISA)-independent features determined by Shao and Brooks {[}8{]}, which in turn builds on earlier work in microarchitecture-independent workload characterization.
Hoste and Eeckout {[}9{]} show that although conventional microarchitecture-dependent characteristics are useful in locating performance bottlenecks {[}10{]}, they are misleading when used as a basis on which to differentiate benchmark applications.
Microarchitecture-independent workload characterization and the associated analysis tool, known as MICA, was proposed to collect metrics to characterize an application independent of particular microarchitectural characteristics.
Architecture-dependent characteristics typically include instructions per cycle (IPC) and miss rates -- cache, branch misprediction and translation look-aside buffer (TLB) -- and are collected from hardware performance counter results, typically PAPI.
However, these characteristics fail to distinguish between inherent program behaviour and its mapping to specific hardware features, ignoring critical differences between architectures such as pipeline depth and cache size.
The MICA framework collects independent features including instruction mix, instruction-level parallelism (ILP), register traffic, working-set size, data stream strides and branch predictability.
These feature results are collected using the Pin {[}12{]} binary instrumentation tool.
In total 47 microarchitecture-independent metrics are used to characterize an application code.
To simplify analysis and understanding of the data, the authors combine principal component analysis with a genetic algorithm to select eight metrics which account for approximately 80\% of the variance in the data set.

A caveat in the MICA approach is that the results presented are not ISA-independent nor independent from differences in compilers.
Additionally, since the metrics collected rely heavily on Pin instrumentation, characterization of multi-threaded workloads or accelerators are not supported.
As such, it is unsuited to conventional supercomputing workloads which make heavy use of parallelism and accelerators.

Shao and Brooks {[}8{]} have since extended the generality of the MICA to be ISA independent.
The primary motivation for this work was in evaluating the suitability of benchmark suites when targeted on general purpose accelerator platforms.
The proposed framework briefly evaluates eleven SPEC benchmarks and examines 5 ISA-independent features/metrics.
Namely, number of opcodes (e.g., add, mul), the value of branch entropy -- a measure of the randomness of branch behaviour, the value of memory entropy -- a metric based on the lack of memory locality when examining accesses, the unique number of static instructions, and the unique number of data addresses.

Related to the paper, Shao also presents a proof of concept implementation (WIICA) which uses an LLVM IR Trace Profiler to generate an execution trace, from which a python script collects the ISA independent metrics.
Any results gleaned from WIICA are easily reproducible, the execution trace is generated by manually selecting regions of code built from the LLVM IR Trace Profiler.
Unfortunately, use of the tool is non-trivial given the complexity of the toolchain and the nature of dependencies (LLVM 3.4 and Clang 3.4).
Additionally, WIICA operates on \texttt{C} and \texttt{C++} code, which cannot be executed directly on any accelerator device aside from the CPU.
Our work extends this implementation to the broader OpenCL setting to collect architecture independent metrics from a hardware-agnostic language -- OpenCL.
Additional metrics, such as Instructions To Barrier (ITB), Vectorization (SIMD) indicators and Instructions Per Operand (SIMT) were also determined and added by us in order to perform a similar analysis for concurrent and accelerator workloads.

The branch entropy measure used by Shao and Brooks {[}8{]} was initially proposed by Yokota {[}13{]} and uses Shannon's information entropy to determine a score of Branch History Entropy.
De Pestel, Eyerman and Eeckhout {[}14{]} proposed an alternative metric, average linear branch entropy metric, to allow accurate prediction of miss rates across a range of branch predictors.
As their metric is more suitable for architecture-independent studies, we adopt it for this work.

Caparrós Cabezas and Stanley-Marbell {[}15{]} present a framework for characterizing instruction- and thread-level parallelism, thread parallelism, and data movement, based on cross-compilation to a MIPS-IV simulator of an ideal machine with perfect caches and branch prediction and unlimited functional units.
Instruction-level and thread-level parallelism are identified through analysis of data dependencies between instructions and basic blocks.
The current version of AIWC does not perform dependency analysis for characterizing parallelism, however, we hope to include such metrics in future versions.

In contrast to our multidimensional workload characterization, models such as Roofline {[}16{]} and Execution-Cache-Memory {[}17{]} seek to characterize an application based on one or two limiting factors such as memory bandwidth.
The advantage of these approaches is the simplicity of analysis and interpretation.
We view these models as capturing a `principal component' of a more complex performance space; we claim that by allowing the capture of additional dimensions, AIWC supports performance prediction for a greater range of applications.

\section{Metrics}\label{metrics}

For each OpenCL kernel invocation, the Oclgrind simulator \textbf{AIWC} tool collects a set of metrics, which are listed in Table \ref{tbl:aiwc-metrics}.

\iftoggle{ACM-BUILD}{
\begin{table*}[t]
\caption{Metrics collected by the \textbf{AIWC} tool ordered by type. \label{tbl:aiwc-metrics}}
}{}
\iftoggle{IEEE-BUILD}{
\begin{table*}[t]
\caption{Metrics collected by the \textbf{AIWC} tool ordered by type. \label{tbl:aiwc-metrics}}
}{}
\iftoggle{LNCS-BUILD}{
\begin{table}[tb]
\caption{Metrics collected by the \textbf{AIWC} tool ordered by type. \label{tbl:aiwc-metrics}}
\begin{adjustbox}{max width=\textwidth}
}{}

\centering

\begin{tabular}{@{}cll@{}}
\toprule

{Type} & {Metric} & {Description}\\\hline

Compute & Opcode & total \# of unique opcodes required to cover 90\% of dynamic
instructions\\
Compute & Total Instruction Count & total \# of instructions executed\\
Parallelism & Work-items & total \# of work-items or threads executed\\
Parallelism & Total Barriers Hit & total \# of barrier instructions\\
Parallelism & Min ITB & minimum \# of instructions executed until a barrier\\
Parallelism & Max ITB & maximum \# of instructions executed until a barrier\\
Parallelism & Median ITB & median \# of instructions executed until a barrier\\
Parallelism & Min IPT & minimum \# of instructions executed per thread\\
Parallelism & Max IPT & maximum \# of instructions executed per thread\\
Parallelism & Median IPT & median \# of instructions executed per thread\\
Parallelism & Max SIMD Width & maximum \# of data items operated on during an instruction\\
Parallelism & Mean SIMD Width & mean \# of data items operated on during an instruction\\
Parallelism & SD SIMD Width & standard deviation across \# of data items affected\\
Memory & Total Memory Footprint & total \# of unique memory addresses accessed\\
Memory & 90\% Memory Footprint & \# of unique memory addresses that cover 90\% of memory accesses\\
Memory & Unique Reads & total \# of unique memory addresses read\\
Memory & Unique Writes & total \# of unique memory addresses written\\
Memory & Unique Read/Write Ratio & indication of workload being (unique reads / unique writes) \\
Memory & Total Reads & total \# of memory addresses read\\
Memory & Total Writes & total \# of memory addresses written\\
Memory & Reread Ratio & indication of memory reuse for reads (unique reads/total reads)\\
Memory & Rewrite Ratio & indication of memory reuse for writes (unique writes/total writes)\\
Memory & Global Memory Address Entropy & measure of the randomness of memory addresses\\
Memory & Local Memory Address Entropy & measure of the spatial locality of memory addresses\\
Control & Total Unique Branch Instructions & total \# of unique branch instructions\\
Control & 90\% Branch Instructions & \# of unique branch instructions that cover 90\%
of branch instructions\\
Control & Yokota Branch Entropy & branch history entropy using Shannon's information entropy\\
Control & Average Linear Branch Entropy & branch history entropy score using the
average linear branch entropy\\
\hline
\end{tabular}

\iftoggle{ACM-BUILD}{
\end{table*}
}{}
\iftoggle{IEEE-BUILD}{
\end{table*}
}{}
\iftoggle{LNCS-BUILD}{
\end{adjustbox}
\end{table}
}{}

The \textbf{Opcode}, \textbf{total memory footprint} and \textbf{90\% memory footprint} measures are simple counts.
Likewise, \textbf{total instruction count} is the number of instructions achieved during a kernel execution.
The \textbf{global memory address entropy} is a positive real number that corresponds to the randomness of memory addresses accessed.
The \textbf{local memory address entropy} is computed as 10 separate values according to increasing number of Least Significant Bits (LSB), or low order bits, omitted in the calculation.
The number of bits skipped ranges from 1 to 10, and a steeper drop in entropy with increasing number of bits indicates greater spatial locality in the address stream.

Both \textbf{unique branch instructions} and the associated \textbf{90\% branch instructions} are counts indicating the count of logical control flow branches encountered during kernel execution.
\textbf{Yokota branch entropy} ranges between 0 and 1, and offers an indication of a program's predictability as a floating point entropy value. {[}13{]}
The \textbf{average linear branch entropy} metric is proportional to the miss rate in program execution; \(p=0\) for branches always taken or not-taken but \(p=0.5\) for the most unpredictable control flow.
All branch-prediction metrics were computed using a fixed history of 16-element branch strings, each of which is composed of 1-bit branch results (taken/not-taken).

As the OpenCL programming model is targeted at parallel architectures, any workload characterization must consider exploitable parallelism and associated communication and synchronization costs.
We characterize thread-level parallelism (TLP) by the number of \textbf{work-items} executed by each kernel, which indicates the maximum number of threads that can be executed concurrently.

Work-item communication hinders TLP, and in the OpenCL setting, takes the form of either local communication (within a work-group) using local synchronization (barriers) or globally by dividing the kernel and invoking the smaller kernels on the command queue.
Both local and global synchronization can be measured in \textbf{instructions to barrier} (ITB) by performing a running tally of instructions executed per work-item until a barrier is encountered under which the count is saved and resets; this count will naturally include the final (implicit) barrier at the end of the kernel.
\textbf{Min}, \textbf{max} and \textbf{median ITB} are reported to understand synchronization overheads, as a large difference between min and max ITB may indicate an irregular workload.

\textbf{Instructions per thread} (IPT) based metrics are generated by performing a running tally of instructions executed per work-item until completion.
The count is saved and resets.
\textbf{Min}, \textbf{max} and \textbf{median IPT} are reported to understand load imbalance.

To characterize data parallelism, we examine the number and width of vector operands in the generated LLVM IR, reported as \textbf{max SIMD width}, \textbf{mean SIMD width} and standard deviation -- \textbf{SD SIMD width}.
Further characterisation of parallelism is presented in the \textbf{work-items} and \textbf{total barriers hit} metrics.

Some of the other metrics are highly dependent on workload scale, so \textbf{work-items} may be used to normalize between different scales.
For example, \textbf{total memory footprint} can be divided by \textbf{work-items} to give the total memory footprint per work-item, which indicates the memory required per processing element.

Finally, unique verses absolute reads and writes can indicate shared and local memory reuse between work-items within a work-group, and globally, which shows the predictability of a workload.
To present these characteristics the \textbf{unique reads}, \textbf{unique writes}, \textbf{unique read/write ratio}, \textbf{total reads}, \textbf{total writes}, \textbf{reread ratio}, \textbf{rewrite ratio} metrics are proposed.
The \textbf{unique read/write ratio} shows that the workload is balanced, read intensive or write intensive.
They are computed by storing read and write memory accesses separately and are later combined, to compute the \textbf{global memory address entropy} and \textbf{local memory address entropy} scores.

\section{Methodology -- Workload Characterization by tooling Oclgrind}\label{methodology-workload-characterization-by-tooling-oclgrind}

AIWC verifies the architecture independent metrics since they are collected on a toolchain and in a language actively executed on a wide range of accelerators -- the OpenCL runtime supports execution on CPU, GPU, DSP, FPGA, MIC and ASIC hardware architectures.
The intermediate representation of the OpenCL kernel code is a subset of LLVM IR known as SPIR -- Standard Portable Intermediate Representation.
This IR forms a basis for Oclgrind to perform OpenCL device simulation, which interprets LLVM IR instructions.

Migrating the metrics presented in the ISA-independent workload characterization paper {[}8{]} to the Oclgrind tool offers an accessible, high-accuracy and reproducible method to acquire these AIWC features.
Namely:

\begin{itemize}
\tightlist
\item
  Accessibility: since the Oclgrind OpenCL kernel debugging tool is one of the most adopted OpenCL debugging tools freely available to date, having AIWC metric generation included as an Oclgrind plugin allows rapid workload characterization.
\item
  High-Accuracy: evaluating the low level optimized IR does not suffer from a loss of precision since each instruction is instrumented during its execution in the simulator, unlike with the conventional metrics generated by measuring architecture driven events -- such as PAPI and MICA analysis.
\item
  Reproducibility: each instruction is instrumented by the AIWC tool during execution, there is no variance in the metric results presented between OpenCL kernel runs.
\end{itemize}

The caveat with this approach is the overhead imposed by executing full solution HPC codes on a slower simulator device.
However, since AIWC metrics do not vary between runs, this is still a shorter execution time than the typical number of iterations required to get a reasonable statistical sample when compared to a MICA or architecture dependent analysis.

\section{Implementation}\label{implementation}

AIWC is implemented as a plugin for Oclgrind, which simulates kernel execution on an ideal compute device.
OpenCL kernels are executed in series, and Oclgrind generates notification events which AIWC handles to populate data structures for each workload metric.
Once each kernel has completed execution, AIWC performs statistical summaries of the collected metrics by examining these data structures.

The \textbf{Opcode} diversity metric updates a counter on an unordered map during each \texttt{workItemBegin} event, the type of operation is determined by examining the opcode name using the LLVM Instruction API.

The number of \textbf{work-items} is computed by incrementing a global counter -- accessible by all work-item threads -- once a \texttt{workItemBegin} notification event occurs.

TLP metrics require barrier events to be instrumented within each thread.
Instructions To Barrier \textbf{ITB} metrics require each thread to increment a local counter once every \texttt{instructionExecuted} has occurred, this counter is added to a vector and reset once the work-item encounters a barrier.
The \textbf{Total Barriers Hit} counter also increments on the same condition.
Work-items are executed sequentially within all work-items in a work-group.
If a barrier is hit the queue moves onto all other available work-items in a ready state.
Collection of the metrics post barrier resumes during the \texttt{workItemClearBarrier} event.

ILP \textbf{SIMD} metrics examine the size of the result variable provided from the \texttt{instructionExecuted} notification, the width is then added to a vector for the statistics to be computed once the kernel execution has completed.

\textbf{Total Memory Footprint} \textbf{90\% Memory Footprint} and Local Memory Address Entropy \textbf{LMAE} metrics require the address accessed to be stored during kernel execution and occurs during the \texttt{memoryLoad}, \texttt{memoryStore}, \texttt{memoryAtomicLoad} and \texttt{memoryAtomicStore} notifications.

Branch entropy measurements require a check during \texttt{instructionExecuted} event on whether the instruction is a branch instruction, if so a flag indicating a branch operation has occurred is set and both LLVM IR labels -- which correspond to branch targets -- are recorded.
On the next \texttt{instructionExecuted} the flag is queried and reset while the current instruction label is compared against which of the two targets were taken, the result is stored in the branch history trace.
The implementation of this is shown in Listing \ref{lst:instructionExecuted}.
Note the \texttt{instructionExecuted} callback is propagated from Oclgrind during every OpenCL kernel instruction -- emulated in LLVM IR.
This function also updates variables to track instruction diversity by counting the occurrences of each instruction, instructions to barrier and other parallelism metrics by running a counter until a barrier is hit, finally, the vectorization -- as part of the parallelism metrics -- are updated by recording the width of executed instructions.
The \texttt{m\_state} variable is shared between all work-items in a work-group and these are stored into a global set of variables using a mutex lock once the work-group has completed execution.

The branch metrics are then computed by evaluating the full history of combined branch's taken and not-taken.

\begin{lstlisting}[float=*t,language=C++, caption={The Instruction Executed callback function collects specific program metrics and adds them to a history trace for later analysis.},label={lst:instructionExecuted}]
void WorkloadCharacterisation::instructionExecuted(...,  const llvm::Instruction *instruction, ...){
    unsigned opcode = instruction->getOpcode();
    std::string opcode_name = llvm::Instruction::getOpcodeName(opcode);
    //update key-value pair of instruction name and its occurrence in the kernel
    (*m_state.computeOps)[opcode_name]++;
    std::string Str = "";
    //if a conditional branch which has labels, store the labels to track
    //in the next instruction which of the two lines we end up in
    if (opcode == llvm::Instruction::Br && instruction->getNumOperands() == 3){
        if(instruction->getOperand(1)->getType()->isLabelTy() &&
                instruction->getOperand(2)->getType()->isLabelTy()){
            m_state.previous_instruction_is_branch = true;
            llvm::raw_string_ostream OS(Str);
            instruction->getOperand(1)->printAsOperand(OS,false);
            m_state.target1 = Str;
            Str = "";
            instruction->getOperand(2)->printAsOperand(OS,false);
            m_state.target2 = Str;
            llvm::DebugLoc loc = instruction->getDebugLoc();
            m_state.branch_loc = loc.getLine();
         }
    }
    //if the last instruction was a branch, log which of the two targets were taken
    else if (m_state.previous_instruction_is_branch == true){
        llvm::raw_string_ostream OS(Str);
        instruction->getParent()->printAsOperand(OS,false);
        if(Str == m_state.target1)
            (*m_state.branchOps)[m_state.branch_loc].push_back(true);//taken
        else if(Str == m_state.target2){
            (*m_state.branchOps)[m_state.branch_loc].push_back(false);//not taken
        }
        m_state.previous_instruction_is_branch = false;
    }
    //counter for instructions to barrier and other parallelism metrics
    m_state.instruction_count++;
    m_state.workitem_instruction_count++;
    //SIMD instruction width metrics use the following
    m_state.instructionWidth->push_back(result.num);
\end{lstlisting}

The \textbf{Total Unique Branch Instructions} is a count of the absolute number of unique locations that branching occurred, while the \textbf{90\% Branch Instructions} indicates the number of unique branch locations that cover 90\% of all branches.
\textbf{Yokota} from Shao {[}8{]}, and \textbf{Average Linear Branch Entropy}, from De Pestel {[}14{]} and have been computed and are also presented based on this implementation.
\texttt{workGroupComplete} events trigger the collection of the intermediate work-item and work-group counter variables to be added to the global suite, while \texttt{workGroupBegin} events reset all the local/intermediate counters.

Finally, \texttt{kernelBegin} initializes the global counters and \texttt{kernelEnd} triggers the generation and presentation of all the statistics listed in Table \ref{tbl:aiwc-metrics}.
The source code is available at the GitHub Repository {[}18{]}.

\section{Demonstration}\label{demonstration}

We now demonstrate the use of \textbf{AIWC} on several scientific application kernels selected from the Extended OpenDwarfs Benchmark Suite {[}5{]}.
These benchmarks were extracted from and are representative of general scientific application codes.
Our selection is not intended to be exhaustive, rather, it is meant to illustrate how key properties of the codes are reflected in the metrics collected by \textbf{AIWC}.

AIWC is run on full application codes, but it is difficult to present an entire summary due to the nature of OpenCL.
Computationally intensive kernels are simply selected regions of the full application codes and are invoked separately for device execution.
As such, the AIWC metrics can either be shown per kernel run on a device, or as the summation of all metrics for a kernel for a full application at a given problem size -- we chose the latter.
Additionally, given the number of kernels presented we believe AIWC will generalize to full codes in other domains.

We present metrics for 11 different application codes -- which includes 37 kernels in total.
Each code was run with four different problem sizes, called \textbf{tiny}, \textbf{small}, \textbf{medium} and \textbf{large} in the Extended OpenDwarfs Benchmark Suite; these correspond respectively to problems that would fit in the L1, L2 and L3 cache or main memory of a typical current-generation CPU architecture.
As simulation within Oclgrind is deterministic, all results presented are for a single run for each combination of code and problem size.

In a cursory breakdown, four selected metrics are presented in Figure \ref{fig:stacked_plots}.
One metric was chosen from each of the main categories, namely, Opcode, Barriers Per Instruction, Global Memory Address Entropy, Branch Entropy (Linear Average).
Each category has also been segmented by colour: blue results represent \emph{compute} metrics, green represent metrics that indicate \emph{parallelism}, yellow represents \emph{memory} metrics and purple bars represent \emph{control} metrics.
Median results are presented for each metric -- while there is no variation between invocations of AIWC, certain kernels are iterated multiple times and over differing domains/data sets.
Each of the 4 sub-figures shows all kernels over the 4 different problem sizes.

For almost all benchmarks the global memory address entropy increases with problem size, whereas the other metrics do not increase.
Notably, memory entropy is low for \texttt{lud\_diagonal}, reflecting memory access with constant strides of diagonal matrix elements, and \texttt{cl\_fdt53Kernel}, again reflecting regular strides generated by downsampling in the discrete wavelet transform.
Note, we do not present \textbf{medium} and \textbf{large} problem sizes for some kernels due to various issues including: a lack of input datasets, failure of AIWC in tracing large numbers of memory and branch operations for entropy calculations.
These issues will be addressed in future work.

Looking at branch entropy, \texttt{bfs\_kernel2} stands out as having by far the greatest entropy.
This kernel is dominated by a single branch instruction based on a flag value which is entirely unpredictable, and could be expected to perform poorly on a SIMT architecture such as a GPU.

Barriers per instruction is quite low for most kernels, with the exception of \texttt{needle\_opencl\_shared\_1} and \texttt{needle\_opencl\_shared\_2} from the Needleman-Wunsch DNA sequence alignment dynamic programming benchmark.
These kernels each have 0.04 barriers per instruction (i.e.~one barrier per 25 instructions), as they follow a highly-synchronized wavefront pattern through the matrix representing matching pairs.
The performance of this kernel on a particular architecture could be expected to be highly dependent on the cost of synchronization.

\begin{figure*}
\centering
\iftoggle{ACM-BUILD}{
%acm
\includegraphics[width=2\columnwidth]{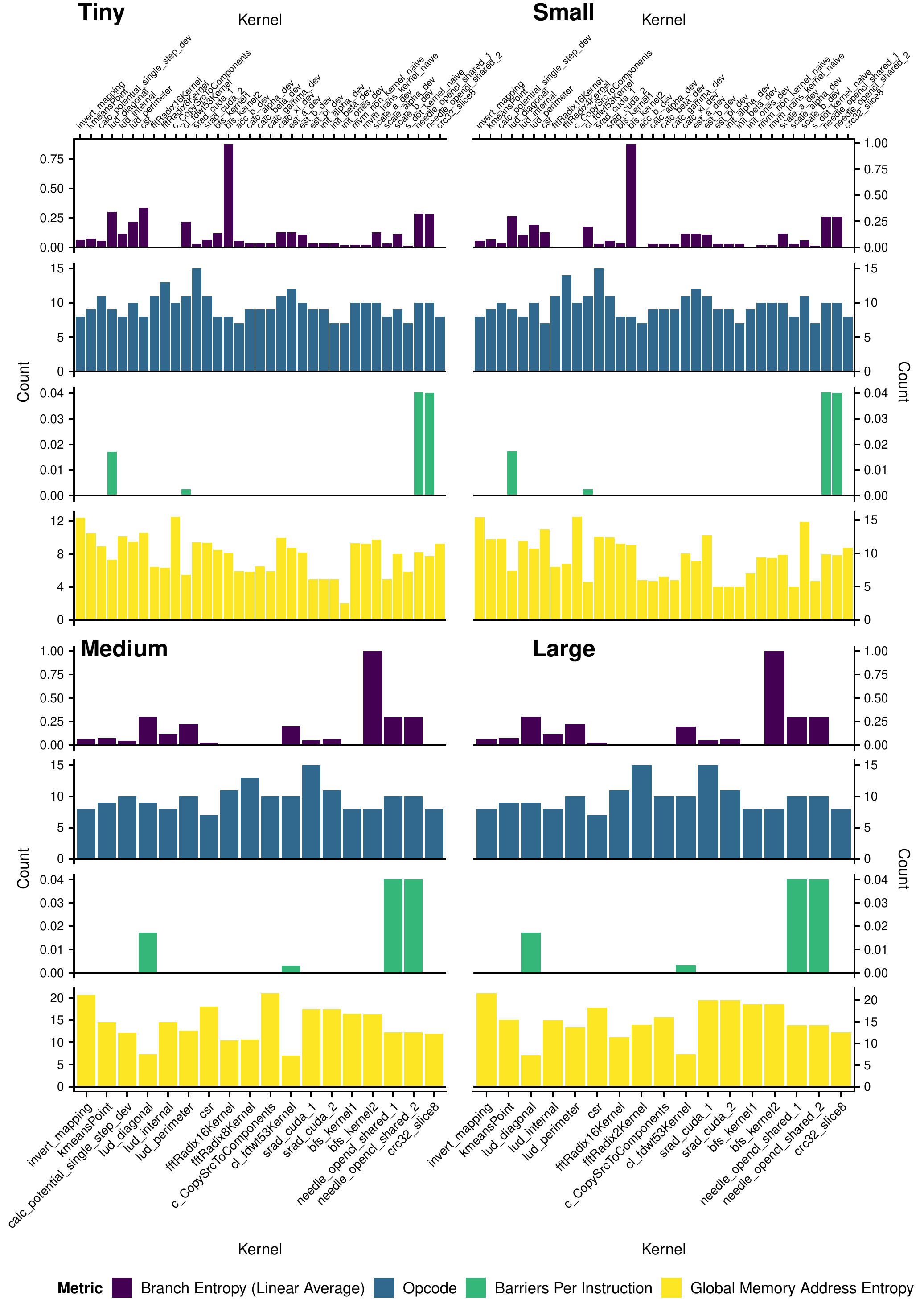}
}{}
\iftoggle{IEEE-BUILD}{
%ieee
\includegraphics[width=1.98\columnwidth]{./figure/draw_stacked_plots-1.pdf}
}{}
\iftoggle{LNCS-BUILD}{
%llncs
\includegraphics[width=0.95\textwidth,height=0.95\textheight,keepaspectratio]{./figure/draw_stacked_plots-1.pdf}
}{}
\caption{Selected AIWC metrics from each category over all kernels and 4 problem sizes.}
\label{fig:stacked_plots} 
\end{figure*}

\begin{figure*}
    \centering
    \newcommand{\plotwidth}{0.66\textwidth}
    \includegraphics[width=\plotwidth]{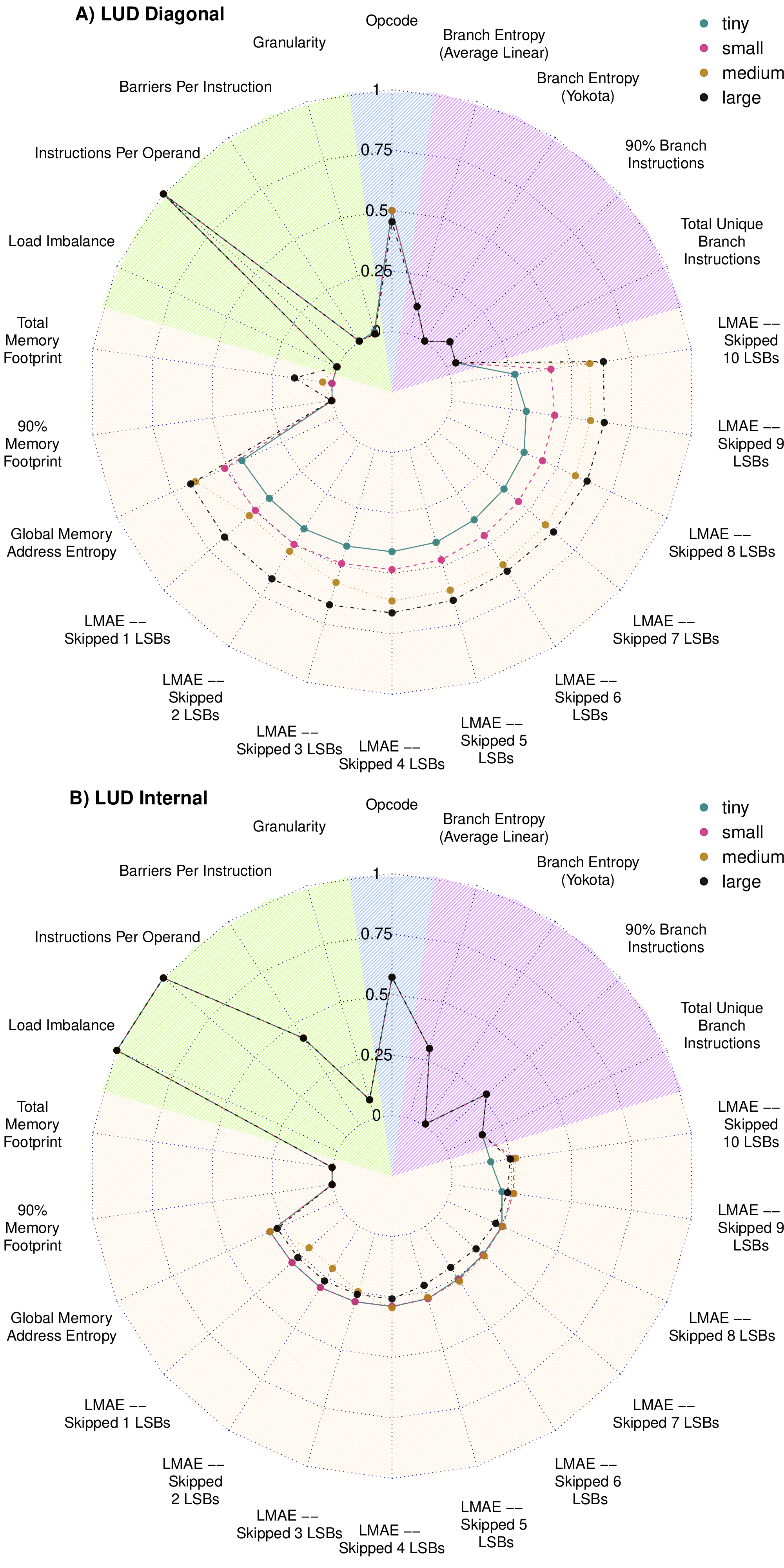}
    \caption{A) and B) show the AIWC features of the \texttt{diagonal} and \texttt{internal} kernels of the LUD application over all problem sizes.}
    \label{fig:kiviat}
\end{figure*}

\begin{figure*}
    \centering
    \newcommand{\plotwidth}{0.66\textwidth}
    \includegraphics[width=\plotwidth]{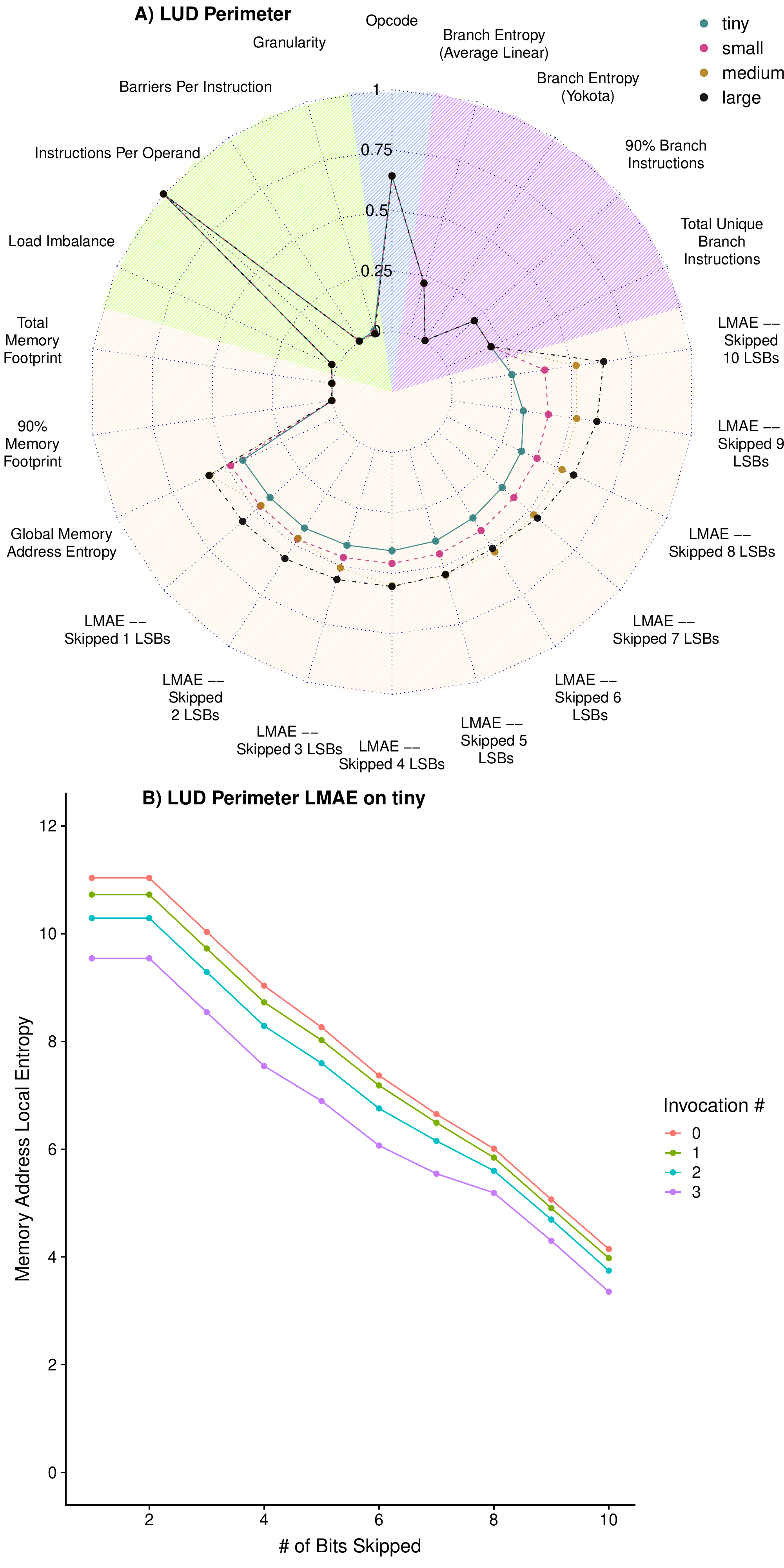}
    \caption{A) shows the AIWC features of the \texttt{perimeter} kernel of the LUD application over all problem sizes. B) shows the corresponding Local Memory Address Entropy for the \texttt{perimeter} kernel over the tiny problem size.}
    \label{fig:kiviat2}
\end{figure*}

\subsection{Detailed Analysis of LU Decomposition Benchmark}\label{detailed-analysis-of-lu-decomposition-benchmark}

We now proceed with a more detailed investigation of one of the benchmarks, \textbf{lud}, which performs decomposition of a matrix into upper and lower triangular matrices.
Following Shao and Brooks {[}8{]}, we present the AIWC metrics for a kernel as a Kiviat or radar diagram, for each of the problem sizes.
Unlike Shao and Brooks, we do not perform any dimensionality reduction but choose to present all collected metrics.
The ordering of the individual spokes is not chosen to reflect any statistical relationship between the metrics, however, they have been grouped into four main categories: green spokes represent metrics that indicate \emph{parallelism}, blue spokes represent \emph{compute} metrics, beige spokes represent \emph{memory} metrics and purple spokes represent \emph{control} metrics.
For clarity of visualization, we do not present the raw AIWC metrics but instead, normalize or invert the metrics to produce a scale from 0 to 1.
The parallelism metrics presented are the inverse values of the metrics collected by AIWC, i.e. \textbf{granularity} = \(1 / \textbf{work-items}\) ; \textbf{barriers per instruction} \(= 1 / \textbf{mean ITB}\) ; \textbf{instructions per operand} \(= 1 / \sum \textbf{SIMD widths}\).

Additionally, a common problem in parallel applications is load imbalance -- or the overhead introduced by unequal work distribution among threads.
A simple measure to quantify imbalance can be achieved using a subset of the existing AIWC metrics and is included as a further derived parallelism metric by computing \textbf{load imbalance} = \textbf{max IPT} \(-\) \textbf{min IPT}.

All other values are normalized according to the maximum value measured across all kernels examined -- and on all problem sizes.
This presentation allows a quick value judgement between kernels, as values closer to the centre (0) generally have lower hardware requirements, for example, smaller entropy scores indicate more regular memory access or branch patterns, requiring less cache or branch predictor hardware; smaller granularity indicates higher exploitable parallelism; smaller barriers per instruction indicates less synchronization; and so on.

The \textbf{lud} benchmark application comprises three major kernels, \textbf{diagonal}, \textbf{internal} and \textbf{perimeter}, corresponding to updates on different parts of the matrix.
The AIWC metrics for each of these kernels are presented -- superimposed over all problem sizes -- in Figure \ref{fig:kiviat} A) B) and Figure \ref{fig:kiviat2} A) respectively.
Comparing the kernels, it is apparent that the diagonal and perimeter kernels have a large number of branch instructions with high branch entropy, whereas the internal kernel has few branch instructions and low entropy.
This is borne out through inspection of the OpenCL source code: the internal kernel is a single loop with fixed bounds, whereas diagonal and perimeter kernels contain doubly-nested loops over triangular bounds and branches which depend on thread id.
Comparing between problem sizes (moving across the page), the large problem size shows higher values than the tiny problem size for all of the memory metrics, with little change in any of the values.

The visual representation provided from the Kiviat diagrams allows the characteristics of OpenCL kernels to be readily assessed and compared.

Finally, we examine the linear memory access entropy (LMAE) presented in the Kiviat diagrams in greater detail.
Figure \ref{fig:kiviat2} B) presents a sample of the linear memory access entropy, in this instance of the LUD Perimeter kernel collected over the tiny problem size.
The kernel is launched 4 separate times during a run of the tiny problem size, this is application specific and in this instance, each successive invocation operates on a smaller data set per iteration.
Note there is a steady decrease in starting entropy, and each successive invocation of the LU Decomposition Perimeter kernel the lowers the starting entropy.
However, the descent in entropy -- which corresponds to more bits being skipped, or bigger the strides or the more localized the memory access -- shows that the memory access patterns are the same regardless of actual problem size.
In general, for cache-sensitive workloads -- such as LU-Decomposition -- a steeper descent between increasing LMAE distances indicates more localized memory accesses, and this corresponds to better cache utilisation when these applications are run on physical OpenCL devices.
It is unsurprising that applications with a smaller working memory footprint would exhibit more cache reuse with highly predictable memory access patterns.

Recently, AIWC has been used for predictive modelling on a set of 15 compute devices including CPUs, GPUs and MIC.
The AIWC metrics generated from the full set of Extended OpenDwarfs kernels were used as input variables in a regression model to predict kernel execution time on each device. {[}19{]}
The model predictions differed from the measured experimental results by an average of 1.1\%, which corresponds to actual execution time mispredictions of 8 \(\mu\)s to 1 second according to problem size.
From the accuracy of these predictions, we can conclude that while our choice of AIWC metrics is not necessarily optimal, they are sufficient to characterize the behaviour of OpenCL kernel codes and identify the optimal execution device for a particular kernel.

\section{Conclusions and Future Work}\label{conclusions-and-future-work}

We have presented the Architecture-Independent Workload Characterization tool (AIWC), which supports the collection of architecture-independent features of OpenCL application kernels.
It is the first workload characterization tool to support multi-threaded or parallel workloads.
These features can be used to predict the most suitable device for a particular kernel, or to determine the limiting factors for performance on a particular device, allowing OpenCL developers to try alternative implementations of a program for the available accelerators -- for instance, by reorganizing branches, eliminating intermediate variables et cetera.
In addition, the architecture independent characteristics of a scientific workload will inform designers and integrators of HPC systems, who must ensure that compute architectures are suitable for the intended workloads.

Caparrós Cabezas and Stanley-Marbell {[}15{]} examine the Berkeley dwarf taxonomy by measuring instruction-level parallelism (ILP), thread-level parallelism (TLP), and data movement.
They propose a sophisticated metric to assess ILP by examining the data dependency graph of the instruction stream.
Similarly, TLP was measured by analysing the block dependency graph.
While we propose alternative metrics to evaluate ILP and TLP -- using the max, mean and standard deviation statistics of SIMD and barrier metrics respectively -- a quantitative evaluation of the dwarf taxonomy using these metrics is left as future work.
We expect that the AIWC metrics will generate a comprehensive feature-space representation which will permit cluster analysis and comparison with the dwarf taxonomy.

We believe AIWC will also be useful in guiding device-specific optimization by providing feedback on how particular optimizations change performance-critical characteristics.
To identify which AIWC characteristics are the best indicators of opportunities for optimization, we are currently looking at how individual characteristics change for a particular code through the application of best-practice optimizations for CPUs and GPUs (as recommended in vendor optimization guides).

A major limitation of running large applications under AIWC is the high memory footprint.
Memory access entropy scores require a full recorded trace of every memory access during a kernel's execution.
However, a graceful degradation in performance is preferable to an abrupt crash in AIWC if virtual memory is exhausted.
For this reason, work is currently being undertaken for an optional build of AIWC with low memory usage by writing these traces to disk.


\begin{thebibliography}{00}

\bibitem{b0}

\hypertarget{ref-spafford2010maestro}{}
 K. Spafford, J. Meredith, and J. Vetter, ``Maestro: Data orchestration and tuning for OpenCL devices,'' \emph{Euro-Par 2010-Parallel Processing}, pp. 275--286, 2010.

\bibitem{b1}

\hypertarget{ref-chaimov2014toward}{}
 N. Chaimov, B. Norris, and A. Malony, ``Toward multi-target autotuning for accelerators,'' in \emph{IEEE international conference on parallel and distributed systems (ICPADS)}, 2014, pp. 534--541.

\bibitem{b2}

\hypertarget{ref-nugteren2015cltune}{}
 C. Nugteren and V. Codreanu, ``CLTune: A generic auto-tuner for OpenCL kernels,'' in \emph{IEEE international symposium on embedded multicore/many-core systems-on-chip (MCSoC)}, 2015, pp. 195--202.

\bibitem{b3}

\hypertarget{ref-price2017analyzing}{}
 J. Price and S. McIntosh-Smith, ``Analyzing and improving performance portability of OpenCL applications via auto-tuning,'' in \emph{Proceedings of the 5th international workshop on OpenCL}, 2017, p. 14.

\bibitem{b4}

\hypertarget{ref-johnston18opendwarfs}{}
 B. Johnston and J. Milthorpe, ``Dwarfs on accelerators: Enhancing OpenCL benchmarking for heterogeneous computing architectures,'' in \emph{Proceedings of the \(47^{th}\) international conference on parallel processing companion}, 2018, pp. 4:1--4:10.

\bibitem{b5}

\hypertarget{ref-price:15}{}
 J. Price and S. McIntosh-Smith, ``Oclgrind: An extensible OpenCL device simulator,'' in \emph{Proceedings of the 3rd international workshop on OpenCL}, 2015, p. 12.

\bibitem{b6}

\hypertarget{ref-kessenich2015}{}
 J. Kessenich, ``A Khronos-Defined Intermediate Language for Native Representation of Graphical Shaders and Compute Kernels.'' 2015.

\bibitem{b7}

\hypertarget{ref-shao2013isa}{}
 Y. S. Shao and D. Brooks, ``ISA-independent workload characterization and its implications for specialized architectures,'' in \emph{IEEE international symposium on performance analysis of systems and software (ISPASS)}, 2013, pp. 245--255.

\bibitem{b8}

\hypertarget{ref-hoste2007microarchitecture}{}
 K. Hoste and L. Eeckhout, ``Microarchitecture-independent workload characterization,'' \emph{IEEE Micro}, vol. 27, no. 3, 2007.

\bibitem{b9}

\hypertarget{ref-ganesan2008performance}{}
 K. Ganesan, L. John, V. Salapura, and J. Sexton, ``A performance counter based workload characterization on Blue Gene/P,'' in \emph{International conference on parallel processing (ICPP)}, 2008, pp. 330--337.

\bibitem{b10}

\hypertarget{ref-prakash2008performance}{}
 T. K. Prakash and L. Peng, ``Performance characterization of SPEC CPU2006 benchmarks on Intel Core 2 Duo processor,'' \emph{ISAST Trans. Comput. Softw. Eng}, vol. 2, no. 1, pp. 36--41, 2008.

\bibitem{b11}

\hypertarget{ref-luk2005pin}{}
 C.-K. Luk \emph{et al.}, ``Pin: Building customized program analysis tools with dynamic instrumentation,'' in \emph{ACM SIGPLAN notices}, 2005, vol. 40, pp. 190--200.

\bibitem{b12}

\hypertarget{ref-yokota2007introducing}{}
 T. Yokota, K. Ootsu, and T. Baba, ``Introducing entropies for representing program behavior and branch predictor performance,'' in \emph{Proceedings of the 2007 workshop on experimental computer science}, 2007, p. 17.

\bibitem{b13}

\hypertarget{ref-depestel2017linear}{}
 S. De Pestel, S. Eyerman, and L. Eeckhout, ``Linear branch entropy: Characterizing and optimizing branch behavior in a micro-architecture independent way,'' \emph{IEEE Transactions on Computers}, vol. 66, no. 3, pp. 458--472, Mar. 2017.

\bibitem{b14}

\hypertarget{ref-CaparrosCabezas:2011:PDM:1989493.1989506}{}
 V. Caparrós Cabezas and P. Stanley-Marbell, ``Parallelism and data movement characterization of contemporary application classes,'' in \emph{Proceedings of the twenty-third annual ACM symposium on parallelism in algorithms and architectures}, 2011, pp. 95--104.

\bibitem{b15}

\hypertarget{ref-williams2009roofline}{}
 S. Williams, A. Waterman, and D. Patterson, ``Roofline: An insightful visual performance model for floating-point programs and multicore architectures,'' \emph{Communications of the Association for Computing Machinery}, 2009.

\bibitem{b16}

\hypertarget{ref-hager2013exploring}{}
 G. Hager, J. Treibig, J. Habich, and G. Wellein, ``Exploring performance and power properties of modern multi-core chips via simple machine models,'' \emph{Concurrency and Computation: Practice and Experience}, vol. 28, no. 2, pp. 189--210, 2013.

\bibitem{b17}

\hypertarget{ref-beau_johnston_2017_1134175}{}
 B. Johnston \emph{et al.}, ``BeauJoh/Oclgrind: Adding AIWC -- An Architecture Independent Workload Characterisation Plugin.'' https://doi.org/10.5281/zenodo.1134175, Dec-2017.

\bibitem{b18}

\hypertarget{ref-johnston18predicting}{}
 B. Johnston, G. Falzon, and J. Milthorpe, ``OpenCL performance prediction using architecture-independent features,'' \emph{International Workshop on High Performance and Dynamic Reconfigurable Systems and Networks (DRSN-2018) (in press)}. http://www.milthorpe.org/pubs/aiwc-perf-prediction, 2018.

\end{thebibliography}
\end{document}